\documentclass[12pt]{article}

\usepackage[a4paper,margin=1in]{geometry}
\usepackage{amsmath,amssymb,amsfonts}
\usepackage{graphicx}
\usepackage{float}
\usepackage{setspace}
\usepackage{hyperref}
\usepackage[numbers]{natbib}
\usepackage{booktabs}
\usepackage{array}
\usepackage{multirow}
\usepackage{enumitem}

\title{Hybrid Classical--Quantum Optimization of Wireless Routing Using QAOA and Quantum Walks}

\author{
Eric Howard$^{1,2}$ \and
Hardique Dasore$^{1}$ \and
Hom Nath Dhungana$^{1}$ \and
Radhika Kuttala$^{1}$ \and
Samuel Murphy$^{3}$ \and
Emma Soo$^{2}$ \and
Shah Haque$^{1}$\\[0.5em]
{\small $^{1}$Southern Cross Institute, Sydney, Australia}\\
{\small $^{2}$Macquarie University, Sydney, Australia}\\
{\small $^{3}$University of Sydney, Sydney, Australia}
}

\date{}

\begin{document}

\maketitle

\begin{abstract}
Routing in wireless communication networks is shaped by mobility, interference, congestion, and competing service requirements, making route selection a high-dimensional constrained optimization problem rather than a simple shortest-path task. This paper investigates the use of hybrid classical--quantum methods for wireless routing, focusing on the Quantum Approximate Optimization Algorithm (QAOA) and quantum walks as candidate mechanisms for exploring complex routing spaces. The paper examines how wireless routing can be expressed as a constrained graph optimization problem in which routing objectives, flow constraints, connectivity requirements, and interference effects are mapped into quantum-compatible Hamiltonian representations. It then discusses how these approaches can be integrated into a hybrid architecture in which classical systems perform network monitoring, graph construction, pre-processing, and deployment, while quantum subroutines are used for selected optimization components. The analysis shows that the potential value of quantum routing lies primarily in the treatment of difficult combinatorial subproblems rather than end-to-end replacement of classical routing frameworks. The paper also highlights practical limitations arising from state preparation, constraint encoding, oracle construction, hardware noise, limited qubit resources, and hybrid execution overhead. It is argued that any meaningful near-term advantage will depend on careful problem decomposition, compact encoding, and tight classical--quantum integration.
\end{abstract}

\section{Introduction}

Wireless communication networks are increasingly defined by volatility, density, and heterogeneity. Routing is no longer a problem of selecting a low-cost path in a relatively stable graph, but a repeated optimization task carried out over environments whose link properties change with mobility, fading, congestion, interference, energy conditions, and service-level requirements. In such systems, route quality depends not merely on connectivity, but on the joint interaction of latency, reliability, energy efficiency, spectral contention, and network-wide state. As wireless infrastructures evolve toward ultra-dense, AI-assisted, and service-adaptive paradigms, the computational burden of high-quality routing rises accordingly.

Classical routing methods remain deeply effective and operationally dominant. Shortest-path methods, distributed routing protocols, metaheuristics, and learning-based decision systems provide practical and well-understood tools for most deployed environments. Yet their limitations become increasingly visible when routing must respond to large search spaces, tight latency constraints, highly coupled interference models, and rapidly changing traffic or topology conditions. The problem becomes especially acute when routing is not a single-objective graph traversal exercise, but a constrained, repeated, multi-objective optimization process over a dynamic network state.

Quantum computation has attracted interest in this context because it introduces a fundamentally different computational model for exploring combinatorial spaces. Instead of sequentially or heuristically evaluating candidate solutions, a quantum system evolves over superposed configurations, enabling amplitude shaping toward desirable outcomes through controlled interference and Hamiltonian dynamics. This does not imply automatic superiority over classical routing methods, nor does it guarantee useful acceleration for arbitrary networking tasks. However, it does create a new framework for posing routing optimization problems, especially those that can be expressed as binary or graph-structured objective functions.

Among the most relevant approaches for near-term quantum hardware is the Quantum Approximate Optimization Algorithm (QAOA), which is explicitly designed for noisy intermediate-scale quantum devices and combines shallow parameterized quantum circuits with classical parameter optimization. Alongside QAOA, quantum search and quantum walks offer distinct mechanisms for path discovery, graph exploration, and search-space acceleration. Grover-style search provides quadratic speedup in unstructured search settings, while quantum walks exploit graph interference structure to enable improved traversal properties under specific topological conditions.

Despite this promise, a realistic assessment must move beyond theoretical speedup claims. In routing, the cost of encoding dynamic classical network data into quantum states, the difficulty of oracle construction, hardware noise, limited qubit counts, coherence limits, and latency introduced by cloud-based quantum access can significantly reduce or eliminate practical advantage. The relevant question is therefore not whether quantum methods can replace classical routing, but under what conditions they can supplement or enhance specific routing subproblems.

This paper develops that argument through a unified discussion of routing formulation, quantum encoding, QAOA, quantum search, quantum walks, and hybrid classical--quantum orchestration. It shows that quantum techniques are most plausible when routing is framed as a constrained optimization task over dynamic graphs, when the optimization kernel can be isolated from the broader network management workflow, and when the quantum routine is tightly integrated with classical monitoring, preprocessing, and decision deployment. The discussion therefore proceeds from routing formulation to quantum encoding, then to QAOA, search and quantum walks, and finally to hybrid architecture, system limitations, and the conditions under which meaningful quantum enhancement may emerge.

\section{Routing Problem Setup and Quantum-Compatible Formulation}

\begin{figure}[H]
    \centering
    \includegraphics[width=0.7\textwidth]{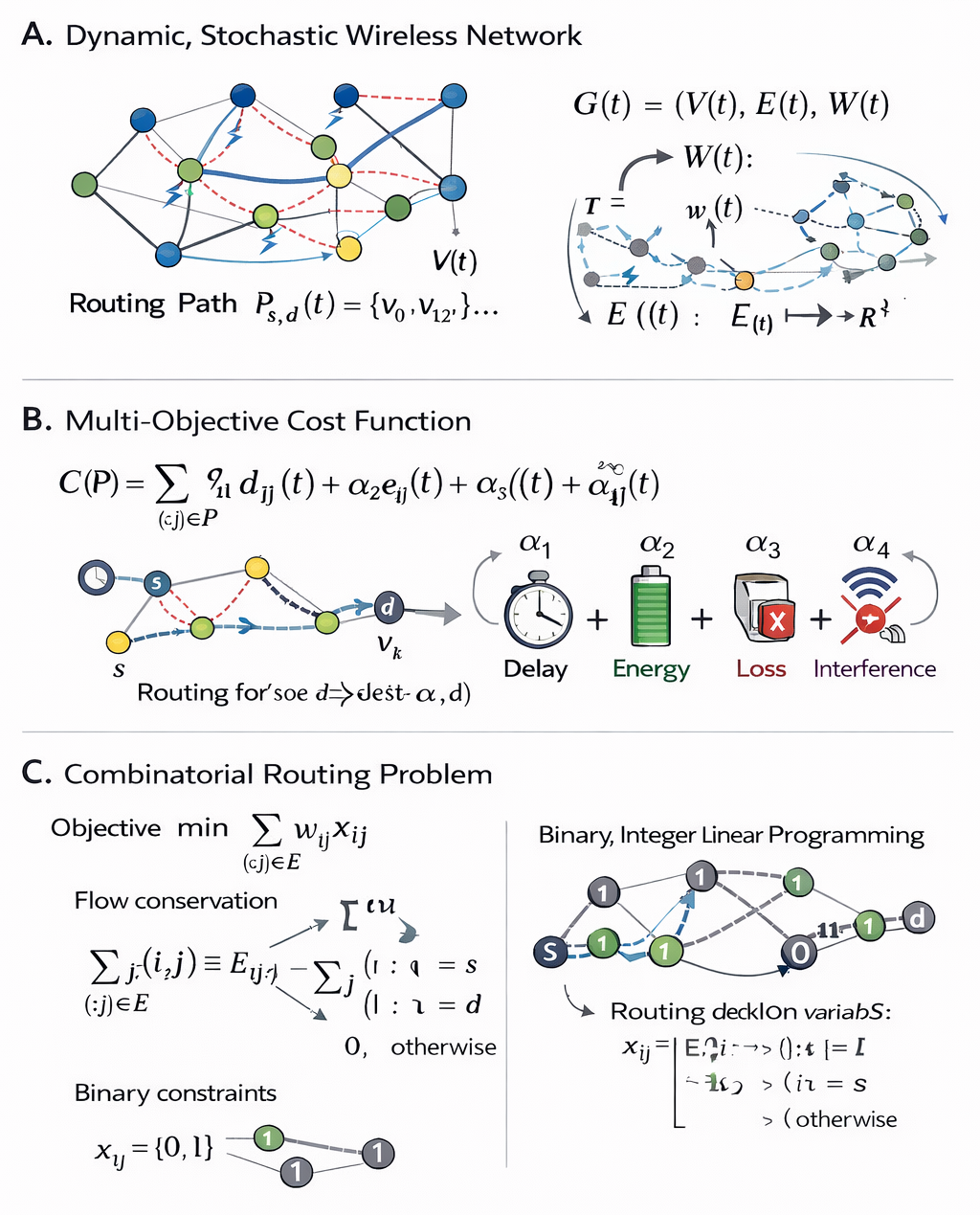}
    \caption{Classical formulation of wireless routing in dynamic networks. The figure illustrates the time-varying graph model, the multi-objective routing cost function, and the combinatorial binary optimization structure with flow conservation constraints that motivate subsequent quantum representations.}
    \label{fig:foundations_routing}
\end{figure}

Before introducing quantum algorithms, it is useful to restate the routing problem in a compact form that makes clear why combinatorial optimization methods are needed. Wireless routing is fundamentally defined on a time-varying graph whose path quality depends not only on connectivity, but also on cumulative cost, feasibility, and dynamic operating conditions. Figure~\ref{fig:foundations_routing} summarizes this classical formulation by showing the evolving graph model, the multi-objective path cost structure, and the constrained binary decision framework that underlies later quantum optimization steps.

Consider a wireless network represented by a time-dependent graph
\[
G(t) = (V(t), E(t), W(t)),
\]
where \(V(t)\) is the node set, \(E(t)\) is the link set, and \(W(t)\) is a time-varying weight assignment over links. Each edge \((i,j)\in E(t)\) may be associated with composite cost terms reflecting delay, energy expenditure, loss probability, congestion, interference, or service-quality penalties. A route between a source node \(s\) and a destination node \(d\) is therefore not simply a path of minimum hop count, but a structured decision that must remain feasible under flow constraints while balancing competing operational priorities. This point is important because contemporary wireless systems demand a richer optimization perspective in which route feasibility, interference coupling, and heterogeneous service requirements all influence the quality of a final routing decision.

A common binary formulation introduces decision variables \(x_{ij}\in\{0,1\}\), where \(x_{ij}=1\) indicates that edge \((i,j)\) is selected. The routing objective may be written as
\[
\min_{x_{ij}} \sum_{(i,j)\in E} w_{ij}x_{ij},
\]
subject to path-consistency and flow-conservation constraints. A simple single-commodity constraint form is
\[
\sum_{j:(i,j)\in E}x_{ij}-\sum_{j:(j,i)\in E}x_{ji}
=
\begin{cases}
1, & i=s,\\
-1, & i=d,\\
0, & \text{otherwise}.
\end{cases}
\]
These constraints ensure that the source emits one unit of flow, the destination absorbs one unit, and all intermediate nodes preserve continuity. From an optimization viewpoint, this transforms routing from a purely descriptive graph process into a constrained combinatorial selection problem. The complexity arises because the number of candidate edge combinations grows rapidly with network size, while only a very small subset correspond to valid, connected, loop-free source-to-destination routes.

When additional terms such as interference or reliability are included, the objective may become multi-objective and non-linear. A representative composite path cost is
\[
C(P)=\sum_{(i,j)\in P}\left(\alpha_1 d_{ij}+\alpha_2 e_{ij}+\alpha_3 \ell_{ij}+\alpha_4 \iota_{ij}\right),
\]
where \(d_{ij}\) denotes delay, \(e_{ij}\) denotes energy cost, \(\ell_{ij}\) denotes loss, and \(\iota_{ij}\) denotes interference contribution. This form highlights why wireless routing is especially difficult in practice. Improving one objective may worsen another. A low-latency route may create higher interference, a highly reliable route may consume more energy, and a congestion-aware route may be dynamically unstable under mobility.

This classical formulation becomes quantum-compatible by mapping binary variables into operators. A standard mapping is
\[
x_{ij}=\frac{1-Z_{ij}}{2},
\]
where \(Z_{ij}\) is a Pauli-\(Z\) operator acting on the qubit assigned to edge \((i,j)\). The routing objective can then be expressed as a cost Hamiltonian
\[
H_C=\sum_{(i,j)\in E} w_{ij}\frac{1-Z_{ij}}{2},
\]
while constraints are embedded through penalty Hamiltonians such as
\[
H_{\text{flow}}=\sum_i\left(\sum_j x_{ij}-\sum_j x_{ji}-b_i\right)^2,
\]
with \(b_i\) encoding the net source-destination demand at node \(i\). The total optimization Hamiltonian becomes
\[
H = H_C + \lambda_1 H_{\text{flow}} + \lambda_2 H_{\text{connect}} + \lambda_3 H_{\text{loop}} + \lambda_4 H_{\text{int}},
\]
where \(H_{\text{connect}}\), \(H_{\text{loop}}\), and \(H_{\text{int}}\) enforce connectivity, loop avoidance, and interference structure, respectively.

This formulation clarifies that routing can be recast not as procedural graph traversal alone, but as constrained Hamiltonian minimization. That reframing enables the use of QAOA, amplitude amplification, and quantum-walk-based graph exploration, all of which rely on the ability to represent routing decisions as structured configurations in Hilbert space.

\section{Quantum Encoding of Wireless Graphs}

The transition from classical routing models to quantum optimization cannot occur directly without an intermediate encoding stage. Before a routing problem can be addressed through QAOA, quantum search, or quantum walks, the graph structure, its weights, and its constraints must be translated into a quantum representation that preserves the essential structure of the original problem while remaining implementable on physical hardware. This encoding stage is therefore not a minor technical detail but one of the most decisive steps in the entire hybrid classical--quantum workflow. A poor encoding can destroy problem structure, inflate resource requirements, and introduce excessive circuit overhead, whereas a well-designed encoding can preserve constraints, support efficient Hamiltonian construction, and enable meaningful exploration of the solution space.

\begin{figure}[H]
    \centering
    \includegraphics[width=0.95\textwidth]{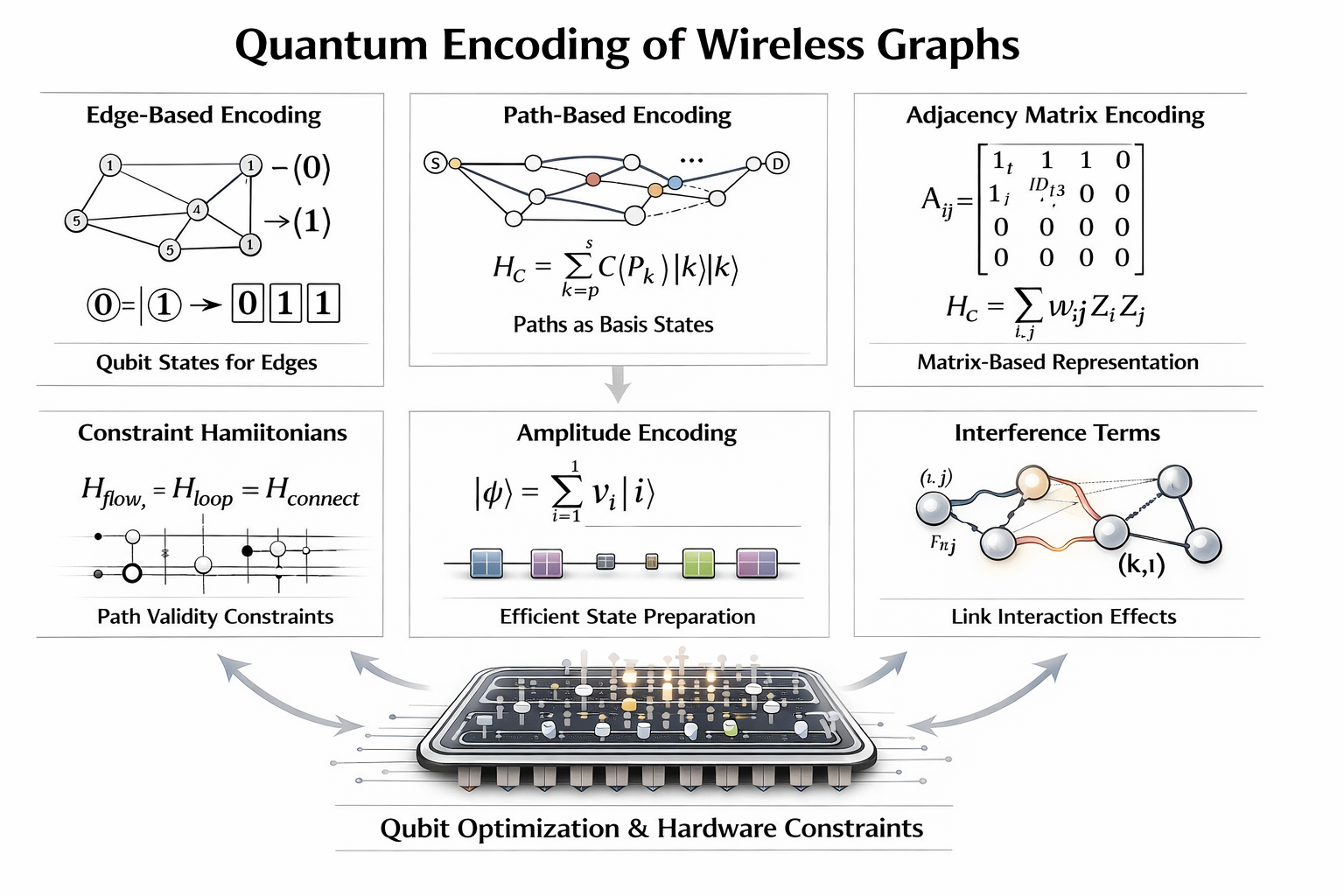}
    \caption{Quantum encoding strategies for wireless graph optimization. The figure presents a conceptual framework for mapping classical wireless network structures into quantum representations suitable for routing optimization. It illustrates edge-based encoding, path-based encoding, adjacency matrix encoding, constraint Hamiltonians, amplitude encoding, interference modelling, and hardware-aware qubit optimization.}
    \label{fig:quantum_wireless_encoding}
\end{figure}

Figure~\ref{fig:quantum_wireless_encoding} presents this intermediate stage explicitly by showing how classical wireless network structures are mapped into quantum representations suitable for routing optimization. The figure captures the main encoding approaches and makes clear that the optimization problem is not sent to the quantum layer as a raw graph. It must first be transformed into a set of qubit variables, interaction terms, and hardware-aware structures that can support routing-related quantum evolution.

As shown in Figure~\ref{fig:quantum_wireless_encoding}, the encoding of wireless graphs involves translating topology, routing constraints, and link interactions into Hamiltonian-compatible forms. One straightforward strategy is edge-based encoding, in which each candidate link is assigned a binary variable and then mapped to a qubit. This method is intuitive and general, and it aligns naturally with Hamiltonian formulations based on edge selection. However, its resource requirements grow with the number of edges, which can become substantial in dense wireless graphs. A second strategy is path-based encoding, in which candidate paths rather than individual edges are encoded as basis states or higher-level variables. This can reduce the burden of enforcing certain structural constraints, but it may scale poorly because the number of candidate paths can become exponentially large.

Adjacency-based and matrix-oriented encodings provide another perspective, especially when the structure of the graph itself is central to the computation. These representations are especially relevant when routing is coupled with graph exploration, spectral analysis, or quantum-walk-based evolution. Amplitude encoding is more qubit-efficient in principle because it can compress large classical structures into fewer qubits, but that advantage is offset by the substantial overhead of preparing the corresponding state. In practical routing systems, the choice of encoding is therefore always a trade-off among qubit count, state preparation cost, constraint enforceability, and hardware compatibility.

Interference modelling is particularly important in wireless environments. Unlike many simplified graph problems, wireless routing cannot assume that edges behave independently. Simultaneous use of certain links may create coupling effects that alter route quality or invalidate apparently good solutions. Encoding schemes must therefore support pairwise or higher-order interaction terms so that interference can be represented as part of the routing Hamiltonian. This requirement complicates implementation because it increases both the algebraic complexity of the optimization model and the physical demands placed on the quantum circuit.

The design of encoding schemes must also account for hardware limitations. The mapping from logical qubits to physical qubits is affected by connectivity constraints, fidelity variations, and coherence times. If two qubits that should interact are not adjacent on the hardware, additional SWAP operations may be required, increasing circuit depth and error accumulation. For this reason, encoding design and hardware-aware optimization are closely linked. A theoretically compact encoding may still be a poor practical choice if it conflicts with the native topology of the available device.

The choice of encoding strategy therefore involves a balance among several competing factors. Edge-based encoding is straightforward but qubit intensive. Path-based encoding may simplify some forms of constraint handling but can explode combinatorially. Amplitude encoding is elegant but costly in state preparation. Adjacency-based encodings are graph-native but may complicate enforcement of routing feasibility. In many cases, hybrid encoding strategies that combine elements of multiple approaches may offer the most practical performance.

This encoding stage forms the conceptual bridge between the routing formulation and the algorithmic methods that follow. Once the graph has been translated into a quantum-compatible representation, QAOA can operate on cost and constraint Hamiltonians, search procedures can act on encoded candidate spaces, and quantum walks can evolve over graph-derived structures. The encoding figure is therefore central to the logic of the paper because it explains how the routing problem actually enters the quantum layer before any optimization is attempted.

\section{QAOA for Wireless Routing}

The Quantum Approximate Optimization Algorithm offers one of the most natural entry points for applying near-term quantum methods to wireless routing. The central appeal of QAOA is that it directly targets constrained combinatorial optimization problems expressed through a cost Hamiltonian, which aligns well with the Hamiltonian formulation of routing introduced above and depends critically on the encoding stage that has just been established. In the variational framework, one prepares a parameterized quantum state and seeks to minimize the expectation value of the routing Hamiltonian over that state. For a trial state \( |\psi(\boldsymbol{\theta})\rangle \), the quantity
\[
\langle \psi(\boldsymbol{\theta})|H|\psi(\boldsymbol{\theta})\rangle
\]
upper-bounds the minimum achievable energy of the routing Hamiltonian. Minimizing this expectation is therefore equivalent to concentrating amplitude on high-quality routing configurations.

The standard QAOA ansatz is written as
\[
|\psi(\boldsymbol{\gamma},\boldsymbol{\beta})\rangle
=
\prod_{l=1}^{p}
e^{-i\beta_l H_M}
e^{-i\gamma_l H_C}
|+\rangle^{\otimes n},
\]
where \(p\) denotes the circuit depth, \(H_C\) is the cost Hamiltonian, and \(H_M\) is the mixer Hamiltonian. The state \( |+\rangle^{\otimes n}\) provides an initial superposition over computational basis states, thereby allowing the algorithm to begin without bias toward any single routing configuration. The cost unitary imprints phase according to routing quality, while the mixer redistributes amplitude through the search space.

In practical routing problems, the choice of mixer is not trivial. A simple transverse-field mixer,
\[
H_M = \sum_i X_i,
\]
is attractive because it is easy to implement. However, it generally does not preserve routing feasibility. The circuit may therefore place amplitude on invalid configurations, requiring strong penalties in the cost Hamiltonian to suppress them. More sophisticated constraint-aware mixers can in principle preserve valid routing structure during evolution, restricting the search to path-consistent subspaces and reducing the burden on penalty terms. Yet these advantages come with increased implementation complexity and often greater circuit depth. This trade-off between hardware simplicity and combinatorial faithfulness is one of the defining issues in applying QAOA to routing.

The routing Hamiltonian itself must encode not only path cost but structural validity. A minimal form,
\[
H = H_C + \lambda H_{\text{flow}},
\]
already combines link-selection cost with flow-conservation penalties, but this is often insufficient for realistic wireless systems. One may also need to penalize disconnected fragments, loops, invalid branchings, or pairwise interference couplings between simultaneously chosen links. For example, interference effects may be represented as
\[
H_{\text{int}} = \sum_{(i,j),(k,l)} \gamma_{ij,kl} x_{ij}x_{kl},
\]
which makes the Hamiltonian sensitive not only to individual edges but also to their interaction structure.

The parameter optimization stage is also crucial. The QAOA objective
\[
F(\boldsymbol{\gamma},\boldsymbol{\beta})
=
\langle \psi(\boldsymbol{\gamma},\boldsymbol{\beta})|H|\psi(\boldsymbol{\gamma},\boldsymbol{\beta})\rangle
\]
is estimated through repeated circuit execution and measurement. A classical optimizer then updates the parameters to reduce this expectation. Methods such as gradient-free optimization or gradient estimation via the parameter-shift rule,
\[
\frac{\partial F}{\partial \theta}
=
\frac{1}{2}
\left[
F\left(\theta+\frac{\pi}{2}\right)
-
F\left(\theta-\frac{\pi}{2}\right)
\right],
\]
allow the classical layer to navigate the parameter landscape. After optimization, the circuit is sampled repeatedly, and each measurement outcome corresponds to a candidate routing configuration. Because measurement is probabilistic, the final output is not a single deterministic route but a distribution over candidate solutions. Practical deployment therefore usually includes classical post-processing, feasibility checking, and route refinement.

\section{Quantum Search for Routing Acceleration}

Quantum search provides a different computational strategy from variational optimization. Instead of shaping a distribution through Hamiltonian minimization, Grover-style algorithms amplify marked solutions within a search space. For a space of size \(N\), classical unstructured search requires \(O(N)\) evaluations, while Grover search reduces this to \(O(\sqrt{N})\). In routing, this is relevant because candidate path sets often grow rapidly with network size.

Given a uniform initial superposition
\[
|\psi_0\rangle = \frac{1}{\sqrt{N}}\sum_x |x\rangle,
\]
a Grover iteration applies an oracle \(O_f\) that marks desirable states and a diffusion operator \(D\) that amplifies them:
\[
O_f|x\rangle =
\begin{cases}
-|x\rangle, & \text{if } x \text{ satisfies the routing criterion},\\
|x\rangle, & \text{otherwise},
\end{cases}
\qquad
D = 2|\psi_0\rangle\langle\psi_0| - I.
\]
The repeated application
\[
G = D O_f
\]
rotates amplitude toward marked states.

In routing, the oracle might identify paths below a threshold cost or configurations satisfying a target latency-reliability trade-off. A quantum minimum-finding variant can then iteratively improve a current candidate route by repeatedly searching for lower-cost alternatives. In principle, this can reduce the number of route evaluations in very large candidate sets. However, the practical bottleneck is oracle construction. Routing oracles are not trivial label functions: they must encode path validity, cost thresholds, and possibly interference or flow constraints. As the routing criterion becomes more realistic, oracle complexity increases, which can substantially offset the quadratic speedup.

Quantum search is therefore best interpreted as a tool for selected routing subproblems rather than a universal replacement for graph algorithms. Its strongest role may lie in threshold-based search, route refinement, and situations where a reduced set of candidates can be searched more efficiently once classical preprocessing has already narrowed the problem space.

\section{Quantum Walks for Wireless Routing}

Quantum walks deserve a separate treatment because they represent a conceptually different approach to routing from both QAOA and Grover-style search. Whereas QAOA treats routing as energy minimization over a combinatorial configuration space, and search-based methods treat routing as the identification of marked candidates, quantum walks treat the network itself as the substrate of evolution. The graph is not merely encoded into a cost function or an oracle. It becomes the object over which quantum amplitude propagates. This distinction is important because routing is fundamentally a graph problem, and quantum walks provide a framework that engages directly with graph topology, connectivity, and structural pathways.

In a continuous-time quantum walk, the state evolves according to
\[
|\psi(t)\rangle = e^{-iHt}|\psi(0)\rangle,
\]
where \(H\) is typically chosen from the adjacency matrix, graph Laplacian, or another matrix representation of network structure. The choice of Hamiltonian determines how probability amplitude spreads over the graph. Unlike a classical random walk, which evolves through stochastic transitions and accumulates probability in a diffusive manner, a quantum walk evolves coherently. Interference between paths can enhance some regions of the graph and suppress others. This means that graph exploration is no longer governed only by local transition probabilities but also by phase structure and global interference patterns.

A routing system must locate feasible trajectories through a changing graph, often under pressure from congestion, interference, or service-class differentiation. Classical graph traversal methods typically examine these possibilities through deterministic computation or randomized heuristics. A quantum walk offers a different possibility: the network can be explored through coherent evolution, potentially revealing promising corridors or target nodes with behaviour that differs substantially from classical diffusion. In some graph families, this can lead to more favourable hitting times or different concentration behaviour around relevant vertices.

If the graph Hamiltonian is modified to reflect edge weights, one can in principle bias the evolution according to cost, reliability, or interference structure. In a weighted environment, desirable links may contribute differently to phase evolution than undesirable ones, allowing the walk to encode more than simple connectivity. The extent to which this can be done efficiently on near-term hardware remains an open practical question, but conceptually it shows that quantum walks are compatible with routing problems that extend beyond unweighted graph traversal.

\section{Hybrid Classical--Quantum Routing Architectures}

\begin{figure}[H]
    \centering
    \includegraphics[width=0.92\textwidth]{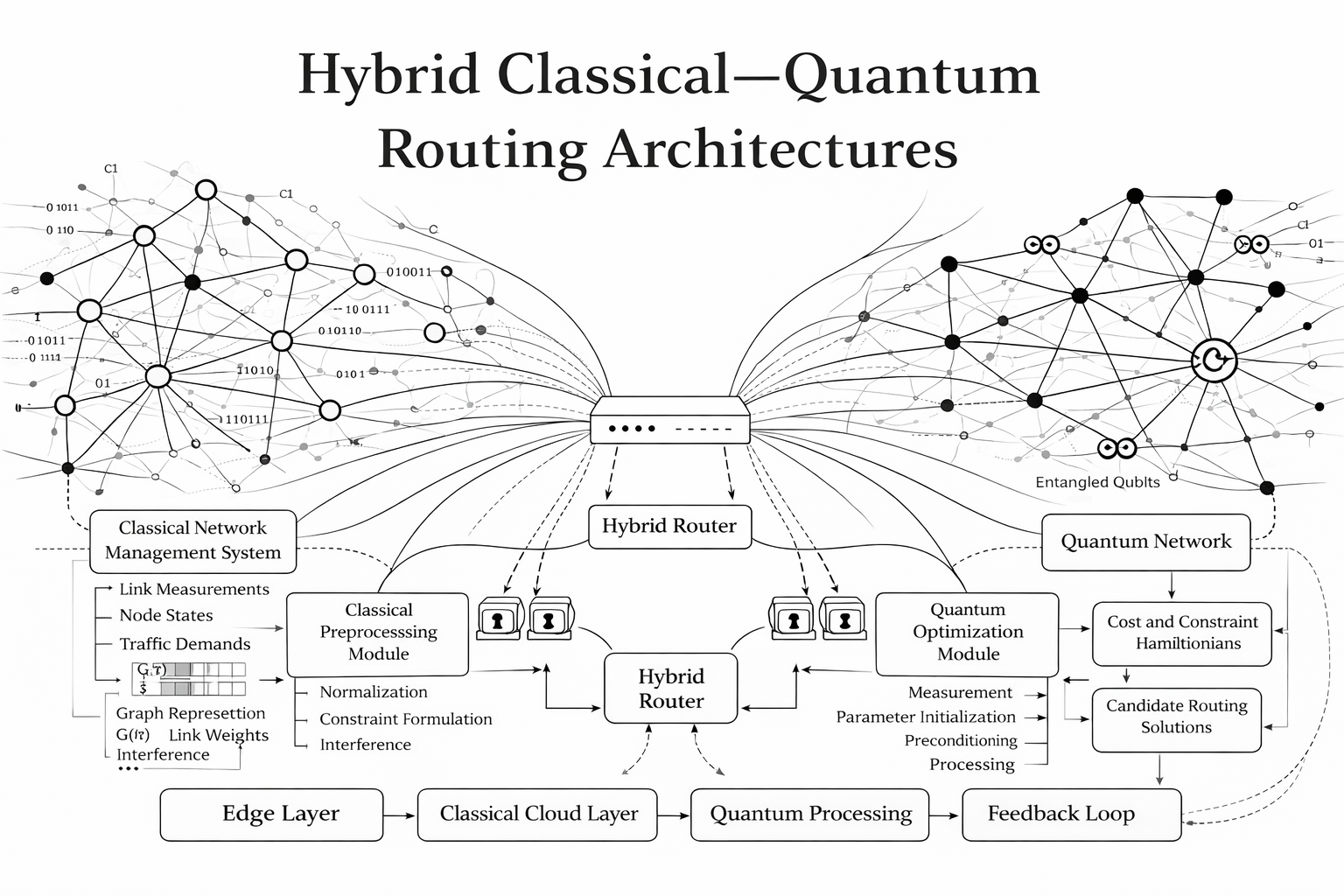}
    \caption{Hybrid classical--quantum routing architecture illustrating the integration of a classical network management and preprocessing pipeline with a quantum optimization pipeline through a central hybrid router. The figure highlights how link measurements, node states, traffic demands, graph representation, and interference-related constraints from the classical domain are transformed into routing optimization inputs, while quantum processing is used to evaluate cost and constraint Hamiltonians, generate candidate routing solutions, and return optimized decisions through a feedback loop.}
    \label{fig:hybrid_arch}
\end{figure}

The practical value of quantum routing depends not only on algorithm design but also on how quantum optimization is embedded within a larger operational workflow. In realistic deployments, classical infrastructure remains responsible for network monitoring, graph construction, constraint preprocessing, orchestration, and final route enforcement, while the quantum layer is invoked selectively for difficult optimization subproblems. Figure~\ref{fig:hybrid_arch} illustrates this layered integration and shows how classical measurements and graph descriptors are transformed into quantum optimization inputs before candidate routing decisions are returned through an iterative feedback loop.

Current quantum hardware remains constrained by qubit counts, gate error rates, limited native connectivity, decoherence, and the absence of fault tolerance at practical routing scales. Wireless routing, meanwhile, depends on continuous data ingestion, rapid state updates, and protocol integration with existing network infrastructure.
Quantum processors are then used selectively for high-complexity combinatorial subproblems such as route optimization over a reduced candidate space, multi-objective refinement, topology reconfiguration search, or graph-exploratory routines based on quantum walks. The architecture succeeds only if the boundary between classical and quantum responsibilities is carefully designed.

At the edge layer, sensors, access points, routers, or controllers gather measurements such as link quality, traffic demand, interference indicators, and mobility state. The classical preprocessing layer converts these measurements into a graph instance \(G(t)\), constructs or updates routing weights, reduces dimensionality where necessary, and builds the optimization Hamiltonian or graph representation for quantum execution.

The quantum layer then executes QAOA, search, or quantum-walk-based routines on the compressed or transformed problem instance. Measurement outputs are returned to the classical layer, which checks feasibility, resolves ambiguity, refines candidate routes, and deploys selected decisions into the live network. Quantum processors are often remotely accessed through centralized cloud infrastructure. This suggests an edge--cloud quantum model in which local systems perform low-latency monitoring and preprocessing, while quantum acceleration is reserved for optimization tasks whose computational burden justifies offloading.

If the routing problem changes faster than quantum optimization can be requested, executed, and reintegrated, any theoretical computational gain becomes operationally irrelevant. Consequently, hybrid architectures are most compelling for semi-static or periodically optimized tasks, topology planning, batch route refinement, or prediction-assisted routing, rather than for every packet-level routing decision. If the quantum backend is unavailable, too noisy, or too slow, the classical system can continue operating using conventional routing heuristics. This resilience is particularly important in networking, where continuity of service is often more important than theoretical optimality.

\section{Complexity, Limitations, and Conditions for Quantum Advantage}

The strongest claims for quantum routing usually rest on asymptotic speedups. Grover search yields \(O(\sqrt{N})\) search in unstructured spaces, quantum walks can outperform classical graph traversal under appropriate conditions, and QAOA may provide improved approximation behaviour in difficult combinatorial landscapes. Yet these claims concern algorithmic kernels, not full routing systems. In practice, routing is an end-to-end process involving monitoring, model construction, encoding, computation, measurement, validation, and deployment.

End-to-end performance must include encoding time, communication overhead, repeated sampling, parameter optimization, feasibility validation, and hardware noise. If the classical routing pipeline requires time \(T_{\text{classical}}\), and the quantum subroutine provides a speedup factor \(S\) but incurs encoding overhead \(T_{\text{encode}}\), the benefit exists only if
\[
T_{\text{encode}} + \frac{T_{\text{classical}}}{S} < T_{\text{classical}}.
\]
In dynamic wireless settings, this condition may be difficult to satisfy because graph states, weights, and interference conditions may need continual updating.

Current devices remain firmly in the NISQ regime. If \(\rho\) denotes the system density matrix, a simple depolarizing model is
\[
\rho \rightarrow (1-p)\rho + \frac{p}{2^n}I,
\]
where \(p\) is the error probability. As circuit depth increases, accumulated noise degrades the quality of samples and weakens the relation between the measured distribution and the intended routing objective. Variational methods are also exposed to barren plateau effects, in which the gradient magnitude becomes vanishingly small as system size and ansatz complexity grow.

In routing, the computational object is rarely an unstructured search space. It is instead a constrained family of feasible paths embedded in a graph with weighted, time-varying, and often correlated edge attributes. If a quantum procedure is to exploit this structure, the graph, the objective, and the admissibility conditions must all be translated into unitary operations, penalty terms, or measurement-interpretable observables. Let \(n_q\) denote the number of logical qubits required for the encoded routing instance, \(d_U\) the effective circuit depth needed to implement the relevant unitaries, and \(N_s\) the number of samples needed to estimate the objective within tolerance \(\varepsilon\). Then the effective runtime burden is not captured by a single asymptotic exponent, but by a composite resource expression of the form
\[
T_{\mathrm{hybrid}}
\sim
T_{\mathrm{prep}} + T_{\mathrm{map}} + N_s d_U \tau_{\mathrm{gate}} + T_{\mathrm{class-opt}} + T_{\mathrm{post}},
\]
where \(T_{\mathrm{prep}}\) denotes classical state preparation and problem assembly, \(T_{\mathrm{map}}\) denotes embedding and compilation overhead, \(\tau_{\mathrm{gate}}\) is the characteristic gate time, \(T_{\mathrm{class-opt}}\) is the classical variational update cost, and \(T_{\mathrm{post}}\) is the route validation and reintegration cost. This decomposition makes clear that even if a quantum subroutine has favourable scaling in one internal component, the total routing pipeline may remain dominated by surrounding stages whose complexity does not inherit the same advantage.

Many routing problems do not require exact global optima; they require high-quality feasible solutions delivered reliably within bounded latency. Variational algorithms such as QAOA naturally operate in this approximate regime, but their performance must be evaluated against strong classical heuristics rather than against exhaustive search alone. If \(C^\star\) denotes the optimal routing cost and \(\hat{C}\) the expected cost of the measured hybrid solution, then one useful metric is the approximation ratio
\[
r = \frac{\hat{C}}{C^\star},
\]
or, in maximization variants, its reciprocal form. However, even this does not fully capture operational value, because routing quality may depend on feasibility rate, robustness under perturbation, and consistency across repeated executions.

Another limitation concerns the spectral and geometric structure of the optimization landscape induced by routing Hamiltonians. Penalty-based encodings frequently generate energy landscapes with narrow feasible basins separated by large infeasible plateaus, especially when flow conservation, loop suppression, interference penalties, and service constraints are all imposed simultaneously. In such cases, increasing the penalty coefficients \(\lambda_k\) does not simply sharpen feasibility; it can also distort the optimization geometry by creating a poorly conditioned landscape in which relevant low-energy states are difficult to resolve under finite sampling noise. Formally, if
\[
H = H_C + \sum_k \lambda_k H_k,
\]
then excessively large \(\lambda_k\) may force the spectrum to be dominated by constraint separation rather than by meaningful discrimination among feasible routes. The consequence is that the variational procedure may spend much of its effort distinguishing infeasible from barely feasible states while failing to rank high-quality feasible paths with sufficient precision. The problem should be large enough that naive classical enumeration is intractable, but sufficiently structured that a compact quantum encoding preserves the dominant constraints without excessive ancillary overhead.

Algorithmic advantage refers to improved scaling of a quantum subroutine relative to a particular classical procedure under an idealized problem model. Systems advantage refers to benefit after embedding, compilation, sampling, and classical orchestration are included. Operational advantage refers to improvement in a live networking context, where solution quality, robustness, reactivity, and integration with control protocols all matter simultaneously.

Many routing tasks contain heterogeneous computational components: graph acquisition, local filtering, candidate reduction, feasibility screening, combinatorial refinement, and route deployment. Only some of these are plausible candidates for quantum acceleration. Let the total routing workload be decomposed as
\[
\mathcal{R} = \mathcal{R}_{\mathrm{monitor}} + \mathcal{R}_{\mathrm{reduce}} + \mathcal{R}_{\mathrm{opt}} + \mathcal{R}_{\mathrm{validate}} + \mathcal{R}_{\mathrm{deploy}},
\]
where \(\mathcal{R}_{\mathrm{opt}}\) is the genuinely hard combinatorial core. Then quantum advantage is meaningful only if the fraction of total cost contained in \(\mathcal{R}_{\mathrm{opt}}\) is sufficiently large and sufficiently isolatable. If most of the routing burden remains in monitoring, graph maintenance, and post-decision reconciliation, accelerating the optimization kernel alone may produce only marginal end-to-end gains. This observation reinforces the hybrid thesis of the paper: the realistic promise of quantum routing lies not in replacing the full stack, but in selectively accelerating the narrowest but hardest combinatorial segments of the workflow.

The routing or topology problem should have a very large combinatorial search space, classical heuristics should struggle with the specific constraint structure, approximate rather than exact solutions should be acceptable, the optimization task should be isolatable and reducible before quantum execution, the hardware should have enough qubits and low enough noise to support the chosen encoding, and the decision latency tolerance should be compatible with hybrid execution overhead.
A hybrid system may improve route diversity, provide better starting points for classical refinement, offer stronger exploration of constrained subspaces, or improve selected multi-objective trade-offs. Such gains may not amount to dramatic asymptotic superiority, but they can still be operationally meaningful in carefully chosen wireless optimization workflows.

\section{Conclusion}

Quantum methods provide a compelling computational framework for routing because wireless routing can be expressed as a high-dimensional, constrained, graph-structured optimization problem whose complexity grows rapidly with network scale, dynamism, and objective coupling. QAOA offers a natural variational framework for Hamiltonian-based route optimization, while amplitude-amplified search and quantum walks provide alternative mechanisms for candidate discovery and graph exploration.

Routing is not only an optimization kernel; it is a real-time systems problem involving state estimation, protocol integration, feasibility handling, latency constraints, and continuous graph updates. Once encoding overhead, oracle complexity, noise, decoherence, hardware access latency, and classical post-processing are included, the practical pathway becomes much narrower than purely asymptotic analysis might suggest.
Quantum routing should not be viewed as an immediate replacement for classical networking methods, nor should it be reduced to a purely speculative abstraction. Its strongest near-term role lies in hybrid architectures in which quantum processors are invoked selectively for carefully structured subproblems, while classical systems continue to dominate monitoring, orchestration, and deployment. Under such conditions, hybrid advantage may emerge before broad quantum advantage.

As quantum hardware matures, the relevance of these methods is likely to increase, especially in dense, adaptive, AI-native communication environments. The long-term significance of quantum routing may therefore lie not in reproducing classical routing protocols with exotic hardware, but in reshaping how difficult network optimization problems are formulated, decomposed, and solved. In that sense, the most important contribution of hybrid classical--quantum routing may be methodological as much as computational.

\bibliographystyle{plainnat}

\end{document}